\documentclass[12pt]{iopart}
\usepackage{graphicx}

\begin{document}

\title{A model for alignment between microscopic rods and vorticity}

\author{M. Wilkinson and H. R. Kennard}

\address{Department of Mathematics and Statistics, The Open University, Milton Keynes, MK7 6AA, England.}

\begin{abstract}

Numerical simulations show that microscopic rod-like bodies suspended 
in a turbulent flow tend to align with the vorticity vector, rather than with
the dominant eignevector of the strain-rate tensor. This paper 
investigates an analytically solvable limit of a model for alignment in a random
velocity field with isotropic statistics. The vorticity varies 
very slowly and the isotropic random flow is equivalent to a 
pure strain with statistics which are axisymmetric about the direction
of the vorticity. We analyse the alignment in a weakly fluctuating 
uniaxial strain field, as a function of the product of the strain relaxation
time $\tau_{\rm s}$ and the angular velocity $\omega$ about the vorticity axis. 
We find that when $\omega\tau_{\rm s}\gg 1$, the rods are predominantly
either perpendicular or parallel to the vorticity.


\end{abstract}


\section{Introduction}
\label{sec: 1}

Microscopic rod-like bodies suspended in a fluid flow rotate in response to 
the velocity gradient of the flow. This introduces a degree of order in the 
orientation of a suspension of particles which can influence its optical or 
rheological properties. The equation of motion for the orientation of
microscopic ellipsoidal particles was obtained by Jeffery \cite{Jef22}. 
The implications of this equation of motion for the orientation have been
considered by numerous authors: for example \cite{Bre62} discusses the 
motion of general axisymmetric particles, \cite{Hin+72} considers the role
of Brownian motion, \cite{Wil+11,Wil+09} discuss the alignment fields 
in (respectively) regular and chaotic flows, and \cite{Par+11,Mis+12} are recent 
experimental contributions which give an extensive list of references.  
There are, however, still aspects which are not thoroughly 
understood. One surprising observation (based upon direct numerical simulation
(DNS) studies of Navier-Stokes turbulence) is that in isotropic fully-developed 
turbulence, rod-like particles show significant alignment with  the vorticity
vector, but negligible alignment with the principal strain axis \cite{Pum+11}. 
This was given a qualitative explanation in \cite{Pum+11}, but it is desirable
to have a model for this surprising effect which can be analysed quantitatively.
 
This paper considers an exactly solvable model for the alignment of rods 
with vorticity. The formulation of this model was motivated by observations
about the velocity gradient field of turbulence. It  has been observed that the fluctuations 
of the vorticity vector decay much more slowly than fluctuations of the rate
of strain: \cite{Pum+11} shows evidence that the correlation functions of strain and 
vorticity both show approximately exponential decay, with decay times 
$\tau_{\rm s}\approx 2.3\tau_{\rm K}$ and 
$\tau_{\rm v}\approx 7.2\tau_{\rm K}$ respectively, where $\tau_{\rm K}$ is the 
Kolmogorov timescale of the turbulence. Similar results were reported earlier by
Girimaji and Pope \cite{Gir+90} and Brunk, Koch and Lion \cite{Bru+98}. 
This observation suggests that it may be helpful to consider the limit
as $\tau_{\rm v}\to \infty$, that is the limit where the vorticity is frozen, in order to
explain the observed alignment. 

We use an Ornstein-Uhlenbeck process to model fluctuations of the velocity gradient,
and consider the limit where the vorticity evolves very
slowly. This model is solved exactly in the limit where the strain which 
occurs over the timescale $\tau_{\rm s}$ is small.
The alignment of the rod direction ${\bf n}$ and the direction of the vorticity 
vector ${\bf e}_\omega$ can be described by computing the probability density function
(PDF) of $z={\bf n}\cdot {\bf e}_\omega$. We find that in these limits the 
 PDF of $z$, denoted by $P(z)$, can be computed exactly. 
This analytically solvable model has a single dimensionless 
parameter, $\zeta\equiv \omega\tau_{\rm s}$, where $\omega$ is the angular velocity of 
rotation about the 
vorticity vector. We find that when $\zeta \gg 1$, the probability density has two sharp
peaks, one at $z=\pm 1$ (indicating perfect alignment with vorticity), 
the other at $z=0$ (implying that the rods are perpendicular to the vorticity). 
In the limit as $\zeta \to \infty$, the peak at $z=\pm 1$ is higher than at $z=0$, but it is 
also narrower, with both peaks containing a finite probability. (Throughout this paper, $\langle X\rangle$ is the expectation value 
of $X$, and we use $P(X)$ to denote its probability density function).

Section \ref{sec: 2} discusses the model which will be solved: the equations 
of motion for a microscopic rod are considered in section \ref{sec: 2.1}, and 
the Ornstein-Uhlenbeck model for the velocity gradient of an isotropic random 
flow is described in section \ref{sec: 2.2}. 
Section \ref{sec: 3} discusses a transformation of the equation
of motion in which the \emph{isotropic} velocity gradient is replaced by a 
pure strain field which is \emph{axisymmetric} about the direction of the 
vorticity vector, and it discusses the parametrisation of such axisymmetric random 
strain fields. Section \ref{sec: 4} considers the general solution for alignment of rod-like 
particles in axisymmetric strain fields, before specialising to the solution of the model 
developed in section \ref{sec: 3}. Section \ref{sec: 5} summarises our conclusions. 
The analysis in section \ref{sec: 4} is closely related to recent work by 
Vincenzi \cite{Vin12}, who analysed the alignment of ellipsoidal particles in an 
axisymmetric Kraichnan-Batchelor model.

\section{Equations of motion}
\label{sec: 2}

\subsection{Non-linear and linear equations of motion for rods}
\label{sec: 2.1}

We consider microscopic objects advected in a fluid with velocity
field $\mbox{\boldmath$u$}(\mbox{\boldmath$r$},t)$.
The objects are assumed to be neutrally buoyant, and smaller than any
lengthscale characterising the fluid, but sufficiently large that
their Brownian motion need not be considered. The motion of the body is described
by the position of its centre, $\mbox{\boldmath$r$}(t)$, and the direction
of a unit vector aligned with its axis, ${\bf n}(t)$. The centre of the
body is assumed to be advected by the fluid flow: 
$\dot {\mbox{\boldmath$r$}}=\mbox{\boldmath$u$}(\mbox{\boldmath$r$},t)$.
The motion of the unit vector ${\bf n}$ defining the axis of symmetry 
is determined by elements of the velocity gradient
tensor, evaluated at the centre of the body:
\begin{equation}
\label{eq: 2.1.1}
A_{ij}(t)=\frac{\partial u_i}{\partial r_j}(\mbox{\boldmath$r$}(t),t)
\end{equation}
where $\mbox{\boldmath$r$}(t)$ is the advected particle trajectory.
The equation of motion of the \emph{director vector} of a microscopic rod-like body is
\cite{Jef22}
\begin{equation}
\label{eq: 2.1.2}
\frac{{\rm d}{\bf n}}{{\rm d}t}={\bf A}(t){\bf n}
-({\bf n}\cdot {\bf A}(t){\bf n}){\bf n}
\ .
\end{equation}
We assume the flow is incompressible, so that $\sum_{i=1}^3 A_{ii}=0$. This tensor
can be decomposed into a symmetric part ${\bf S}$, which is termed the strain rate,
and an antisymmetric part $\mbox{\boldmath$\Omega$}$, which is the vorticity
tensor:
\begin{equation}
\label{eq: 2.1.3}
{\bf A}={\bf S}+\mbox{\boldmath$\Omega$}
\ ,\ \ \
{\bf S}^{\rm T}={\bf S}
\ ,\ \ \
{\mbox{\boldmath$\Omega$}}^{\rm T}=-\mbox{\boldmath$\Omega$}
\ .
\end{equation}

If the velocity gradient matrix were constant in time, the equation of motion (\ref{eq: 2.1.2}) 
would imply that the vector ${\bf n}$ would become aligned with the eigenvector 
corresponding to its largest eigenvalue. However, numerical simulations of equation 
(\ref{eq: 2.1.2}) for velocity fields of fully developed turbulence show 
a different, and unexpected, phenomenon \cite{Pum+11}. It is found that the direction vector ${\bf n}$ has negligible correlation with the dominant strain eigenvector, but that it
does have a quite pronounced correlation with the vorticity vector, $\mbox{\boldmath$\omega$}$. 

Our analysis of the alignment due to the motion (\ref{eq: 2.1.2}) will use 
an observation due to Szeri \cite{Sze93}: the non-linear 
equation (\ref{eq: 2.1.2}) can be solved by considering a companion linear equation
for a vector $\mbox{\boldmath$x$}(t)$, which evolves under the action of a monodromy matrix
${\bf M}(t)$:
\begin{equation}
\label{eq: 2.1.4}
\mbox{\boldmath$x$}(t)={\bf M}(t)\mbox{\boldmath$x$}(0)
\ ,\ \ \ \ 
\frac{{\rm d}{\bf M}}{{\rm d}t}={\bf A}(t)\,{\bf M} 
\end{equation}
where the initial conditions are ${\bf M}(0)={\bf I}$ (the identity matrix) and $\mbox{\boldmath$x$}(0)={\bf n}(0)$.
The solution to (\ref{eq: 2.1.2}) is obtained by normalising the solution of (\ref{eq: 2.1.4}):
\begin{equation}
\label{eq: 2.1.5}
{\bf n}(t)=\frac{\mbox{\boldmath$x$}(t)}{|\mbox{\boldmath$x$}(t)|}
\ .
\end{equation}
The advantage of this approach is that it is easier to solve the linear equation
(\ref{eq: 2.1.4}) than the non-linear equation (\ref{eq: 2.1.2}).

\subsection{Ornstein-Uhlenbeck model for velocity gradients in isotropic flows}
\label{sec: 2.2}

In this section we describe a simple stochastic model for the matrix ${\bf A}(t)$
in isotropic random flows. A version of this model was used by Vincenzi {\sl et al} 
\cite{Vin+07}, and its structure is suggested by the observations in \cite{Bru+98}. 
The model was also considered in \cite{Pum+11}, which gave a detailed account of its implementation. 
Here we give a brief summary. 

It is known that the elements of ${\bf S}$ and $\mbox{\boldmath$\Omega$}$
fluctuate randomly about zero, with different timescales $\tau_{\rm s}$ and $\tau_{\rm v}$
respectively. Their correlation functions are well approximated by exponential
functions. This suggests modelling the elements of ${\bf S}$ and $\mbox{\boldmath$\Omega$}$
by Ornstein-Uhlenbeck processes \cite{Uhl+30,vKa81}. 
The three independent components of the vorticity will be modelled by:
\begin{equation}
\label{eq: 2.2.1}
\dot \Omega_{ij}=-\frac{1}{\tau_{\rm v}}\Omega_{ij}+\sqrt{2\,D_{\rm v}}\eta_{ij}(t)
\end{equation}
where the $\eta_{ij}(t)$ are independent white-noise signals, satisfying
\begin{equation}
\label{eq: 2.2.2}
\langle \eta(t)\rangle=0
\ ,\ \ \
\langle \eta(t)\eta(t')\rangle=\delta (t-t')
\ .
\end{equation}
This model predicts that the correlation function of $\Omega_{ij}$ is exponential \cite{Uhl+30,vKa81}:
\begin{equation}
\label{eq: 2.2.3}
\langle \Omega_{ij}(t_1)\Omega_{ij}(t_2)\rangle=D_{\rm v}\tau_{\rm v}\exp(-|t_1-t_2|/\tau_{\rm v})
\ .
\end{equation}
The components of the strain-rate matrix are generated by a further 
six Ornstein-Uhlenbeck processes, with a different correlation time
$\tau_{\rm s}$. The off-diagonal elements are generated by a process
of the same form as (\ref{eq: 2.2.1}), with the diffusion coefficient in 
(\ref{eq: 2.2.2}) replaced by $D_{\rm s}$. The diagonal elements of the 
strain-rate matrix must satisfy $\sum_{i=1}^3 S_{ii}=0$,
which is the incompressibility condition, $\mbox{\boldmath$\nabla$}\cdot \mbox{\boldmath$u$}=0$.
This constraint is satisfied by the solution of the following Ornstein-Uhlenbeck equations
\begin{equation}
\label{eq: 2.2.4}
\dot S_{ii}=-\frac{1}{\tau_{\rm s}}S_{ii}+\sqrt{2\,D_{\rm d}}\left[\eta_{ii}(t)-\frac{1}{3}\sum_{j=1}^3\eta_{jj}(t)\right]
\ .
\end{equation}
The elements $\Omega_{ij}$ and $S_{ij}$ generated by these processes are statistically
independent, apart from the constraint that $\sum_i S_{ii}=0$. The variances of the off-diagonal,
diagonal and vorticity elements are respectively denoted $\langle S^2_{\rm o}\rangle$,
$\langle S^2_{\rm d}\rangle$ and $\langle \Omega^2\rangle$, and are related to the relaxation
times and diffusion rates by $\langle S_{\rm o}^2\rangle=D_{\rm s}\tau_{\rm s}$, 
$\langle S_{\rm d}^2\rangle=\frac{2}{3}D_{\rm d}\tau_{\rm s}$ and 
$\langle \Omega^2 \rangle=D_{\rm v}\tau_{\rm v}$. 
The requirement that the statistics of the model are invariant under rotations (so that
it describes a velocity gradient with isotropic statistics) gives $D_{\rm d}=2D_{\rm s}$, 
so that this model has four parameters: $\tau_{\rm s}$, $\tau_{\rm v}$, $D_{\rm s}$
and $D_{\rm v}$. Note that the diffusion coefficients have dimension $[D]=T^{-3}$,
implying that the model has three independent dimensionless parameters.
In the following we consider the limit as $\tau_{\rm v}/\tau_{\rm s}\to \infty$, so that 
the vorticity is frozen, with angular velocity $\omega$. We also assume that
$D_{\rm s}\tau_{\rm s}^3\ll 1$, so that the strain fluctuations are small. 
This leaves one dimensionless parameter, which we will take to be $\zeta=\omega\tau_{\rm s}$.

\section{Transformation to an axisymmetric pure strain model}
\label{sec: 3}

\subsection{The frozen vorticity limit}
\label{sec: 3.1}

In this section we consider the alignment of rod-like particles in an isotropic flow, where there 
is a non-zero vorticity which is slowly varying. 
The approach is to transform the equation of motion to a reference 
frame which rotates around the axis of vorticity. In this coordinate system, the strain
field oscillates in directions which are perpendicular to the vorticity
vector, in addition to having random temporal fluctuations. 
The effect of these oscillations is to reduce the effective intensity of the 
random strain field in directions perpendicular to the vorticity vector, so that
an isotropic problem with vorticity is transformed to an axisymmetric model
with a velocity gradient which is a pure strain. This reduction was also
discussed in \cite{Pum+11}, but is included here for the convenience of the reader.
  
In order to isolate the effect of the vorticity in the equation of motion
for the monodromy matrix, ${\bf M}$, we introduce another
monodromy matrix ${\bf M}_0$ which evolves under the vorticity alone:
\begin{equation}
\label{eq: 3.1.1}
\dot{\bf M}=({\bf S}+\mbox{\boldmath$\Omega$}){\bf M}
\ ,\ \ \
\dot{\bf M}_0=\mbox{\boldmath$\Omega$}\,{\bf M}_0
\ .
\end{equation}
Note that ${\bf M}_0(t)$ is just a rotation matrix, describing rotation
about an axis in the direction of the vorticity vector $\mbox{\boldmath$\omega$}$. 
The two monodromy matrices may be related by writing
\begin{equation}
\label{eq: 3.1.2}
{\bf M}(t)={\bf M}_0(t) {\bf K}(t)
\end{equation}
where ${\bf K}(t)$ is an evolution matrix which describes the effect 
of the shear. An elementary calculation shows that ${\bf K}$ has the equation
of motion
\begin{equation}
\label{eq: 3.1.3}
\dot {\bf K}=\mbox{\boldmath$\sigma$}(t){\bf K}
\end{equation}
where
\begin{equation}
\label{eq: 3.1.4}
\mbox{\boldmath$\sigma$}={\bf M}_0^{-1}{\bf S}{\bf M}_0
\end{equation}
is obtained from ${\bf S}$ by applying a time-dependent rotation. 
Consider the form of the matrix $\mbox{\boldmath$\sigma$}$.
In the case where the vorticity vector is frozen, and equal
to $\mbox{\boldmath$\Omega$}_0$, the matrix ${\bf M}_0$
is a rotation matrix: ${\bf M}_0=\exp(\mbox{\boldmath$\Omega$}_0\,t)$. 
Without loss of generality we can consider the case where the vorticity is aligned with 
the $z$-axis, with magnitude $\Omega=2\omega$, where $\omega$ is
the rotational angular velocity, so that
${\bf M}_0$ is a rotation matrix of the form
\begin{equation}
\label{eq: 3.1.5}
\fl {\bf M}_0=\exp(\mbox{\boldmath$\Omega$}_0t)={\bf R}(\omega t)=
\left(
\begin{array}{ccc}
\cos \omega t& -\sin \omega t& 0 \cr
\sin \omega t & \cos \omega t& 0 \cr
0                  &      0              & 1
\end{array}
\right)
\equiv
\left(
\begin{array}{ccc}
c & -s & 0 \cr
s & c  & 0 \cr
0 & 0  & 1
\end{array}
\right)
\ .
\end{equation}
If the elements of ${\bf S}$ are $S_{ij}$, the elements of $\mbox{\boldmath$\sigma$}$ are
\begin{equation}
\label{eq: 3.1.6}
\!\!\!\!\!\!\!\!\!\!\!\!\!\!\!\!\!\!\!\!\!\!\!\!\!\!\!\!\!\!\!\!\! 
\mbox{\boldmath$\sigma$}=
\left(
\begin{array}{ccc}
c^2 S_{11}+s^2 S_{22}+2csS_{12}&(c^2-s^2)S_{12}+cs(S_{22}-S_{11})&cS_{13}-sS_{23}\cr
(c^2-s^2)S_{12}+cs(S_{22}-S_{11})&s^2S_{11}+c^2S_{22}-2csS_{12}&cS_{23}+sS_{13}\cr
cS_{13}-sS_{23}&cS_{23}+sS_{13}&S_{33}
\end{array}
\right)
\ .
\end{equation}
Note that all of the off-diagonal components oscillate with angular frequency $\omega$ or
$2\omega$. The diagonal component in the direction of the vorticity vector does not oscillate,
but the other diagonal elements contain both oscillatory terms and non-oscillatory terms.

\subsection{Limit of short correlation time for strain rate}
\label{sec: 3.2}

Now consider the case where the strain rate ${\bf S}$ is sufficiently small 
that the strain which accumulates over its correlation time $\tau_{\rm s}$ is very small.
In this case the evolution of the matrix ${\bf K}$ (defined by equation (\ref{eq: 3.1.3}))
can be described by a diffusive process. Specifically, we consider the evolution of (\ref{eq: 3.1.3}) over a time period $\delta t$ which is large compared to the correlation 
time of the strain field $\tau_{\rm s}$, but still sufficiently small that the strain which
accumulates over this time interval is small. We write
\begin{equation}
\label{eq: 3.2.1}
{\bf K}(t+\delta t)=\left({\bf I}+\delta \mbox{\boldmath$\Sigma$}(\delta t,t)\right)\,{\bf K}(t)
\end{equation}
where the $\delta \mbox{\boldmath$\Sigma$}$ are small and may be assumed to be 
 random matrices, chosen independently at each timestep. We characterise the evolution (\ref{eq: 3.1.3}) by 
computing the statistics of the random strain increments $\delta \mbox{\boldmath$\Sigma$}$,
which are in turn obtained from the random strain ${\bf S}(t)$ using 
(\ref{eq: 3.1.3}) and (\ref{eq: 3.1.6}).
The advantage of considering the small elements $\mbox{\boldmath$\Sigma$}$ is that
they are small random increments which are applied independently at each timestep.
This enables their effect to be analysed using a Fokker-Planck equation.
First consider the relation between the elements of the matrices 
$\delta\mbox{\boldmath$\Sigma$}$ and $\mbox{\boldmath$\sigma$}$. By integrating 
(\ref{eq: 3.1.3}) and using the definition (\ref{eq: 3.2.1}) we obtain
\begin{equation}
\label{eq: 3.2.2}
\delta\mbox{\boldmath$\Sigma$}(\delta t,t)=\int_t^{t+\delta t}{\rm d}t'\ 
\mbox{\boldmath$\sigma$}(t')\,\left({\bf I}+\delta\mbox{\boldmath$\Sigma$}(t'-t,t)\right)
\ .
\end{equation}
Iterating this expression, taking $t=0$, and suppressing the initial time $t$ in the arguments of $\delta \mbox{\boldmath$\Sigma$}$ we obtain
\begin{equation}
\label{eq: 3.2.3}
\delta\mbox{\boldmath$\Sigma$}(\delta t)=\int_0^{\delta t}{\rm d}t_1\ 
\mbox{\boldmath$\sigma$}(t_1)
+\int_0^{\delta t}{\rm d}t_1\int_0^{t_1}{\rm d}t_2\ 
\mbox{\boldmath$\sigma$}(t_1)\mbox{\boldmath$\sigma$}(t_2)
+O(\sigma^3)
\ .
\end{equation}
Using the fact that the correlation time is assumed to satisfy $\delta t\gg \tau_{\rm s}$,
we can write
\begin{equation}
\label{eq: 3.2.4}
\delta \mbox{\boldmath$\Sigma$}(\delta t)=
\int_0^{\delta t}{\rm d}t\ \mbox{\boldmath$\sigma$}(t)+
\frac{\delta t}{2}\int_{-\infty}^\infty {\rm d}t\ \langle \mbox{\boldmath$\sigma$}(t)\mbox{\boldmath$\sigma$}(0)\rangle
+O(\delta t^{3/2})
\ . 
\end{equation}
The first of term is a random variable with mean zero and size $O(\delta t^{1/2})$, 
giving rise to a diffusion term in a Fokker-Planck equation. The second term represents
a drift at a velocity which is well-defined in the limit as $\delta t\to 0$. The remaining
terms may be neglected. In order to formulate the Fokker-Planck equation, we must
determine the statistics of the increments $\delta \Sigma_{ij}(\delta t)$.

If $\omega \tau_{\rm s}\ll 1$, the effect of the oscillatory terms in equation (\ref{eq: 3.1.6})
is negligible. Let us consider how to treat the problem when $\omega\tau_{\rm s}$ is not small.
To simplify the discussion, consider the quantity
\begin{equation}
\label{eq: 3.2.5}
\delta F=\int_0^{\delta t} {\rm d}t\ f(t)\cos (\omega t)
\end{equation}
where $\delta t/\tau_{\rm s} \gg 1$, and where $f(t)$ is a random function
which satisfies
\begin{equation}
\label{eq: 3.2.6}
\langle f(t)\rangle=0
\ ,\ \ \
\langle f(t)f(t')\rangle=C(t-t')
\ .
\end{equation}
The spectral intensity $I(\nu)$ of the fluctuations of $f(t)$ is defined by
\begin{equation}
\label{eq: 3.2.7}
I(\nu)=\int_{-\infty}^\infty {\rm d}t\ \exp({\rm i}\nu t)\, C(t)
\end{equation}
and we shall assume that $C(-t)=C(t)$, so that $I(-\omega)=I(\omega)$.
The expectation value of $\delta F$ is equal to zero. Its variance is
\begin{eqnarray}
\label{eq: 3.2.8}
\langle \delta F^2\rangle&=\int_0^{\delta t}{\rm d}t_1
\int_0^{\delta t}{\rm d}t_2\ \langle f(t_1)f(t_2)\rangle
\cos(\omega t_1)\cos(\omega t_2)
\nonumber \\
&=\frac{1}{2}\int_0^{\delta t}{\rm d}t_1\int_0^{\delta t}{\rm d}t_2\
C(t_1-t_2)[\cos(\omega(t_1-t_2))+\cos(\omega(t_1+t_2))]
\nonumber \\
&=\frac{1}{2}\delta t\int_{-\infty}^\infty {\rm d}s\ C(s)
\cos(\omega s)+O(\delta t^2)
\nonumber \\
&=\frac{1}{4}\delta t [I(\omega)+I(-\omega)]+O(\delta t^2)=\frac{1}{2}\delta t\, I(\omega)+O(\delta t^2)
\ .
\end{eqnarray}
The third steps assumes that $\omega \delta t \gg 1$, as well as $\delta t/\tau \gg 1$.

Now consider the effect of the random strain model defined 
by (\ref{eq: 2.2.1})-(\ref{eq: 2.2.4}) in the limit
where the timescale $\tau_{\rm s}$ of the fluctuations of $S_{ij}(t)$ is very small.
We assume that the functional form of the spectral intensity of each
component $S_{ij}$ is the same, but that their variances are different,
so that the spectral intensity of $S_{ij}(t)$ is $\langle S_{ij}^2\rangle I(\nu)$,
implying that the intensity function is normalised so that $I(0)=1$.
We represent the effect of the randomly fluctuating strain
field described by (\ref{eq: 3.1.6}) by an effective strain field with
diffusive fluctuations. 
Note that $\delta t$ is assumed to satisfy $\delta t/\tau \gg 1$,
despite being \lq small'. By
applying (\ref{eq: 3.2.8}), variance of $\delta \Sigma_{11}$ is
\begin{eqnarray}
\label{eq: 3.2.9}
\fl \langle \delta \Sigma_{11}^2\rangle =\int_0^{\delta t}{\rm d}t_1 \int_0^{\delta t}{\rm d}t_2
\ \biggl \langle \left[\frac{1}{2}(1+\cos 2\omega t_1)S_{11}(t_1)+\frac{1}{2}(1-\cos 2\omega t_2)S_{22}(t_1)+\sin 2\omega t_1 S_{12}(t_1)\right] 
\nonumber \\
\ \quad \times \left[\frac{1}{2}(1+\cos 2\omega t_2) S_{11}(t_2)+\frac{1}{2}(1-\cos 2\omega t_2) S_{22}(t_2)+\sin 2\omega t_2 S_{12}(t_2)\right]\biggr \rangle
\nonumber \\
=\delta t\int_{-\infty}^\infty {\rm d}\tau\ 
\frac{1}{8}[2+\cos 2 \omega \tau]\left\langle S_{11}(\tau) S_{11}(0)\right\rangle
+\frac{1}{8}[2+\cos 2\omega \tau]\left\langle  S_{22}(\tau) S_{22}(0)\right\rangle
\nonumber \\
\quad +\frac{1}{2}\cos(2\omega \tau)\left\langle S_{12}(\tau)S_{12}(0)\right\rangle
+\frac{1}{4}[2-\cos(2\omega \tau)]\langle S_{11}(\tau)S_{22}(0)\rangle+O(\delta t^2)
\nonumber \\
=\frac{\delta t}{8}[2+I(2\omega)]\langle S_{11}^2\rangle
+\frac{\delta t}{8}[2+I(2\omega)]\langle S_{22}^2\rangle
+\frac{\delta t}{2}I(2\omega) \langle S_{12}^2\rangle
\nonumber \\
\quad +\frac{\delta t}{4}[2-I(2\omega)]\langle S_{11}S_{22}\rangle\ +O(\delta t^2)
\end{eqnarray}
Using the same approach, the full set of non-zero covariances of $\delta \Sigma_{ij}$ is
\begin{eqnarray}
\label{eq: 3.2.10}
\langle \delta \Sigma_{11}^2\rangle =\langle \delta \Sigma_{22}^2\rangle 
&=\frac{\delta t}{4}\left[(2+I(2\omega))\langle S_{11}^2\rangle+(2-I(2\omega))\langle  S_{11}S_{22}\rangle+2I(2\omega)\langle S_{12}^2\rangle\right]
\nonumber \\
\langle \delta\Sigma_{11}\delta \Sigma_{22}\rangle
&=\frac{\delta t}{4}\left[(2-I(2\omega))\langle S_{11}^2\rangle+(2+I(2\omega))\langle  S_{11}
S_{22}\rangle -2I(2\omega)\langle S_{12}^2\rangle\right]
\nonumber \\
\langle \delta\Sigma_{12}^2\rangle
&=\frac{\delta t}{4}\left[I(2\omega)\langle S_{11}^2\rangle-I(2\omega)\langle 
S_{11}S_{22}\rangle+2I(2\omega)\langle S_{12}^2\rangle\right]
\nonumber \\
\langle \delta \Sigma_{33}^2\rangle&=\delta t\,I(0)\,\left[\langle 2 S_{11}^2\rangle + 2\langle S_{11}S_{22}\rangle\right]
\nonumber \\
\langle \Sigma_{13}^2\rangle=\langle \Sigma_{23}^2\rangle 
&=\delta t I(\omega) \langle S_{13}^2\rangle
\ .
\end{eqnarray}
Finally, we must consider the mean values of the increments $\delta \Sigma_{ij}(\delta t)$.
As an example, consider the evaluation of $\langle \delta \Sigma_{11}\rangle$. From
the second term in the right hand side of (\ref{eq: 3.2.4}), we have
\begin{eqnarray}
\label{eq: 3.2.11}
\langle \delta \Sigma_{11}\rangle&=\frac{\delta t}{2}\int_{-\infty}^\infty {\rm d}t\ 
\sum_{j=1}^3\langle \sigma_{1j}(t)\sigma_{j1}(0)\rangle 
\nonumber \\
&=\frac{\delta t}{2}\int_{-\infty}^\infty {\rm d}t\ 
c^2\langle S_{11}(t)S_{11}(0)\rangle
+s^2\langle S_{11}(t)S_{22}(0)\rangle
\nonumber \\
&+(c^2-s^2)\langle S_{12}(t)S_{12}(0)\rangle
+c\langle S_{13}(t)S_{13}(0)\rangle
\nonumber \\
&=\frac{\delta t}{4}[
(1+I(2\omega))\langle S_{11}^2\rangle
+(I-I(2\omega))\langle S_{11}S_{22}\rangle
\nonumber \\
&\quad +2I(2\omega)\langle S_{12}^2\rangle
+2I(\omega)\langle S_{13}^2\rangle
]
\ .
\end{eqnarray}
Only the diagonal elements of $\delta \mbox{\boldmath$\Sigma$}$ have a 
non-zero contribution to the mean at $O(\delta t)$: we define velocity 
coefficients $\mu_j$ as follows
\begin{eqnarray}
\label{eq: 3.2.12}
\langle \delta \Sigma_{11}\rangle=\mu_1 \delta t&=
\frac{\delta t}{4}\bigg[
(1+I(2\omega))\langle S_{11}^2\rangle
+(I-I(2\omega))\langle S_{11}S_{22}\rangle
\nonumber \\
&+2I(2\omega)\langle S_{12}^2\rangle
+2I(\omega)\langle S_{13}^2\rangle
\bigg]
\nonumber \\
\langle \delta \Sigma_{22}\rangle=\mu_2\delta t&=\mu_1 \delta t
\nonumber \\
\langle \delta \Sigma_{33}\rangle=\mu_3 \delta t&=\frac{\delta t}{4}\left[
4\langle S_{11}^2\rangle+4\langle S_{11}S_{22}\rangle
+4I(\omega)\langle S_{13}^2\rangle
\right]
\ .
\
\end{eqnarray}
\subsection{Uniaxial random strain in three dimensions}
\label{sec: 3.3}

In sections \ref{sec: 3.1} and \ref{sec: 3.2} we showed how an isotropic model
with frozen vorticity and rapidly fluctuating strain can be represented by an
axisymmetric model where the velocity gradient is a pure strain 
$\mbox{\boldmath$\sigma$}$. In the limit where the strain which occurs over the 
correlation time $\tau_{\rm s}$ is small, the effect of this strain is represented 
by a product of matrices ${\bf I}+\delta\mbox{\boldmath$\Sigma$}$, where 
the small increments $\delta \mbox{\boldmath$\Sigma$}$ are independently 
distributed at each timestep of size $\delta t$. They have diffusive
fluctuations, so that $\delta \mbox{\boldmath$\Sigma$}=O(\delta t^{1/2})$. 
The matrix $\mbox{\boldmath$\sigma$}$ is traceless, representing the fact
that the velocity field is incompressible. The matrix $\delta \mbox{\boldmath$\Sigma$}$
need not, however, satisfy ${\rm tr}(\delta \mbox{\boldmath$\Sigma$})=0$, 
although it is clear that the leading order term in (\ref{eq: 3.2.4}) is traceless.
In this section we discuss how to parametrise such axisymmetric strain fields. 

We take this axis of rotational symmetry to be ${\bf e}_3$; the general case is obtained
from this one by applying rotation matrices. The strain is described by a
$3\times 3$ matrix $\delta \mbox{\boldmath$\Sigma$}$, which we 
write in the form 
\begin{equation}
\label{eq: 3.3.1}
\delta \mbox{\boldmath$\Sigma$}=\left(\begin{array}{ccc}
\delta A & \delta C & \delta D \cr
\delta C & \delta B & \delta E \cr
\delta D &\delta E & -(\delta A+\delta B)
\end{array}\right)
+
\left(\begin{array}{ccc}
\mu_1\delta t &  0 & 0 \cr
0 & \mu_1 \delta t & 0 \cr
0 & 0 & \mu_3 \delta t
\end{array}\right)
\end{equation}
where $\delta A$, $\delta B$, $\delta C$, $\delta D$ and $\delta E$ are random 
variables with mean value zero, and diffusive fluctuations: $\langle \delta A\rangle=0$
and $\langle \delta A^2\rangle=2D_{AA}\delta t$, $\langle \delta A\delta  B\rangle=2D_{AB}\delta t$, etc. 

Applying a rotation about the ${\bf e}_3$ axis by angle
$\theta$ to the random component of $\delta \mbox{\boldmath$\Sigma$}$ 
gives a transformed matrix, with elements
$\delta A'$, $\delta B'$, $\delta C'$, $\delta D'$  and $\delta E'$, given by
\begin{eqnarray}
\label{eq: 3.3.2}
\delta A'&=\cos^2\theta \delta A+\sin^2\theta \delta B+2\cos\theta\sin\theta \delta C
\nonumber \\
\delta B'&=\sin^2\theta \delta A+\cos^2\theta \delta B-2\cos\theta\sin\theta \delta C
\nonumber \\
\delta C'&=(\cos^2\theta-\sin^2\theta)\delta C+\cos\theta\sin\theta(\delta B-\delta A)
\nonumber \\
\delta D'&=\cos\theta \delta D+\sin\theta \delta E
\nonumber \\
\delta E'&=\cos\theta \delta E-\sin\theta \delta D
\
\end{eqnarray}
where $c=\cos \theta$ and $s=\sin\theta$.
The non-random diagonal component is invariant under rotation about ${\bf e}_3$.
Note that $\delta A'+\delta B'=\delta A+\delta B$, so that the element 
$\delta \Sigma_{33}$ is invariant under rotation.

We require that the statistics of the elements are invariant under the rotation angle
$\theta$. It is clear that $\delta A$ and $\delta B$ must have the same variance, as must
$\delta D$ and $\delta E$. Without loss of generality, we can consider a model with $\langle \delta A^2\rangle=2\delta t$.
We therefore characterise the model by the following statistics, where $\alpha$, $\beta$, $\gamma$
are three constants:
\begin{eqnarray}
\label{eq: 3.3.3}
\langle \delta A^2\rangle=\langle \delta B^2\rangle&=2\delta t
\nonumber \\
\langle \delta A \delta B\rangle&=2 \alpha \delta t
\nonumber \\
\langle \delta C^2\rangle&=2 \beta \delta t
\nonumber \\
\langle \delta D^2\rangle = \langle \delta E^2\rangle&=2\gamma \delta t
\ .
\end{eqnarray}
Other covariances, such as $\langle \delta B\delta E\rangle$, are equal to zero. 
The requirement that the statistics of the rotated matrix are independent of $\theta$
leads to the equations
\begin{eqnarray}
\label{eq: 3.3.4}
\langle \delta A'^2\rangle&=2[c^4+s^4+2c^2s^2\alpha+4c^2s^2\beta]\delta t=2\delta t
\nonumber \\
\langle A'B'\rangle &=2[-4c^2s^2\beta+2c^2s^2+(c^4+s^4)\alpha] \delta t=2\alpha \delta t
\nonumber \\
\langle C'^2\rangle&=2[(c^4+s^4-2c^2s^2)\beta+c^2s^2(2-2\alpha)]\delta t=2\beta \delta t
\ .
\end{eqnarray}
Rotational invariance therefore leads to an equation which must the satisfied by $\alpha$ and $\beta$:
\begin{equation}
\label{eq: 3.3.5}
\alpha+2\beta=1
\end{equation}
so that the model for a uniaxial random strain has four independent parameters,
which we can take to be $\alpha$, $\gamma$, $\mu_1$ and $\mu_3$. 

For a special choice of these parameters the model is isotropic. Clearly
this requires $\mu_1=\mu_3$, and  
$\langle \delta C^2\rangle=\langle \delta D^2\rangle=\langle \delta E^2\rangle$, implying  $\gamma=\beta$. Also, requiring
$\langle (\delta A+\delta B)^2\rangle=\langle \delta A^2\rangle=\langle \delta B^2\rangle$ gives
$2+2\alpha=1$. Solving these equations we find that the the covariances of the 
random terms are fixed in the isotropic case 
\begin{equation}
\label{eq: 3.3.6}
\alpha=-\frac{1}{2}
\ ,\ \ \
\beta=\gamma=\frac{3}{4}
\ , \ \ \ 
\mu_3=\mu_1 
\ .
\end{equation}
Another notable limit of the model is the case where the 
matrix is diagonal: this model is $\beta=\gamma=0$, implying
$\alpha=1$. 

\section{Alignment in random strain fields}
\label{sec: 4}

\subsection{General solution in a diffusive axisymmetric strain}
\label{sec: 4.1}

In section \ref{sec: 3} we described the construction of a model for the alignment
of microscopic rods with vorticity, in which the velocity gradient is represented
as a strain field with diffusive fluctuations, axisymmetric
about the direction of the vorticity.
First we consider the alignment of rod-like particles under a 
succession of independent random shears 
${\bf I}+\delta \mbox{\boldmath$\Sigma$}$, which satisfy the conditions
derived in section \ref{sec: 3.3} for the shear statistics to be uniaxial,
before discussing the specific model for rod alignment in section \ref{sec: 4.2}.
 
Using the approach summarised by equations (\ref{eq: 2.1.4}) and (\ref{eq: 2.1.5}), 
the direction vector ${\bf n}$ of a rod-like particle evolves
according to the linear equation
\begin{equation}
\label{eq: 4.1.1}
({\bf I}+\delta \mbox{\boldmath$\Sigma$}){\bf n}(t)=(1+\delta R){\bf n}(t+\delta t)
\end{equation}
where $\delta \mbox{\boldmath$\Sigma$}$ is the infinitesimal strain in time $\delta t$, 
previously introduced in equation (\ref{eq: 3.2.2}), and $\delta R$ is the fractional change
in length of the vector under the linear evolution equation.
Write ${\bf n}(t+\delta t)={\bf n}(t)+\delta {\bf n}+O(\delta {\bf n}^2)$,
where $\delta {\bf n}\cdot {\bf n}=0$. Because of rotational symmetry about the $z$-axis,
we can assume without loss of generality that the $y$ component of ${\bf n}$ is
equal to zero. We therefore consider the following orthogonal basis of unit vectors
\begin{eqnarray}
\label{eq: 4.1.2}
{\bf n}&=(\sin\theta,0,\cos \theta)=(x,0,z)
\nonumber \\
{\bf m}&=(-\cos\theta,0,\sin\theta)=(-z,0,x)
\nonumber \\
{\bf k}&=(0,1,0)
\ .
\end{eqnarray}
where $\theta$ is the polar angle, and $z=\cos\theta$. Writing
$\delta {\bf n}=\delta X{\bf m}+\delta Y{\bf k}$, we have
\begin{equation}
\label{eq: 4.1.3}
{\bf n}(t+\delta t)={\bf n}+\delta X{\bf m}+\delta Y{\bf k}-\frac{1}{2}(\delta X^2+\delta Y^2){\bf n}
+O(\delta {\bf n}^3)
\ .
\end{equation}
By taking the dot product of (\ref{eq: 4.1.1}) in turn with ${\bf n}$, ${\bf m}$ and ${\bf k}$, we find,
respectively to leading order
\begin{equation}
\label{eq: 4.1.4}
\delta R\sim {\bf n}\cdot \delta \mbox{\boldmath$\Sigma$}\,{\bf n}\equiv \delta \Sigma_{nn}
\end{equation}
and
\begin{eqnarray}
\label{eq: 4.1.5}
{\bf m}\cdot \delta {\bf n}(1+\delta R) \ \sim \ {\bf m}\cdot \delta \mbox{\boldmath$\Sigma$}\,{\bf n}\equiv \delta \Sigma_{mn}
\nonumber \\
{\bf k}\cdot \delta {\bf n}(1+\delta R) \ \sim\ {\bf k}\cdot \delta \mbox{\boldmath$\Sigma$}\,{\bf n}\equiv \delta \Sigma_{kn}
\
\end{eqnarray}
Let us characterise the evolution of ${\bf n}$ through the evolution of its
projection onto the ${\bf e}_3$ axis, namely
\begin{equation}
\label{eq: 4.1.6}
z={\bf e}_3\cdot {\bf n}
\ .
\end{equation}
This is a convenient choice because $z$ will have a uniform
probability density function for an isotropic strain field.
Using (\ref{eq: 4.1.3}), we find that
\begin{equation}
\label{eq: 4.1.7}
z+\delta z\equiv {\bf e}_3 \cdot {\bf n}(t+\delta t)=\cos \theta +\sin\theta \delta X -\frac{1}{2}\cos\theta(\delta X^2+\delta Y^2)
\ .
\end{equation}
We define the drift velocity $v_z$ and diffusion coefficient $D_z$ of $z$ by
\begin{equation}
\label{eq: 4.1.8}
\langle \delta z\rangle=v_z \delta t
\ , \ \ \
\langle \delta z^2\rangle=2D_z\delta t
\ .
\end{equation}
Using (\ref{eq: 4.1.9}) and (\ref{eq: 4.1.4}), (\ref{eq: 4.1.5}) we obtain
\begin{equation}
\label{eq: 4.1.9}
v_z \delta t=x\langle \delta \Sigma_{mn}-\delta \Sigma_{nn}\delta \Sigma_{mn}\rangle-\frac{z}{2}\langle \delta S_{mn}^2+\delta \Sigma_{kn}^2\rangle
+O(\delta t^{3/2})
\end{equation}
and
\begin{equation}
\label{eq: 4.1.10}
D_z\delta t=\frac{1}{2}(1-z^2)\langle \delta \Sigma_{mn}^2\rangle+O(\delta t^{3/2})
\ .
\end{equation}
Now consider that statistics of the fluctuations of $z$ for the uniaxial
strain model. For the model defined in section \ref{sec: 3.3}, we have
\begin{eqnarray}
\label{eq: 4.1.11}
\delta \Sigma_{nn}&=\delta Ax^2+2\delta Dxz-(\delta A+\delta B)z^2+\mu_1x^2\delta t+\mu_3z^2\delta t
\nonumber \\
\delta \Sigma_{mn}&=\delta D(x^2-z^2)-(2\delta A+\delta B)xz+(\mu_3-\mu_1)xz\delta t
\nonumber \\
\delta \Sigma_{kn}&=\delta Cx+\delta Ez
\end{eqnarray}
where $x=\sqrt{1-z^2}$. We can combine these relations with (\ref{eq: 4.1.9}) and (\ref{eq: 4.1.10})
to determine $D_z$ and $v_z$:
\begin{eqnarray}
\label{eq: 4.1.12}
\fl D_z\delta t=\frac{1-z^2}{2} \langle [\delta D(1-2z^2)-(2\delta A+\delta B)xz]^2\rangle
\nonumber \\
\fl v_z\delta t=-x\langle [\delta A(1-2z^2)-\delta Bz^2+2\delta Dxz][\delta D(1-2z^2)-(2\delta A+\delta B)xz]\rangle
\nonumber \\
\fl \qquad \qquad -\frac{z}{2}\langle [\delta D(1-2z^2)-(2\delta A+\delta B)xz]^2\rangle-\frac{z}{2}\langle [\delta Cx+\delta Ez]^2\rangle
+\Delta \mu x^2z\delta t
\
\end{eqnarray}
where $\Delta \mu=\mu_3-\mu_1$.
Using the statistics of the elements $\delta A$, $\delta B$, $\delta C$, $\delta D$ and $\delta E$, and ordering
the resulting expressions as polynomials in $z$, we have:
\begin{eqnarray}
\label{eq: 4.1.13}
D_z&=\frac{1}{2}(1-x^2)\left[\gamma+(5+4\alpha-4\gamma)z^2-(5+4\alpha-4\gamma)z^4\right]
\nonumber \\
v_z&=\left(\frac{7}{4}+\frac{5}{4}\alpha-\frac{5}{2}\gamma+\Delta \mu \right)z+\left(-\frac{37}{4}-\frac{29}{4}\alpha+\frac{15}{2}\gamma-\Delta \mu \right)z^3
\nonumber \\
& +\left(\frac{15}{2}+6\alpha-6\gamma\right)z^5
\ .
\end{eqnarray}
The steady-state probability density for $z$, namely $P(z)$, satisfies
\begin{equation}
\label{eq: 4.1.14}
v_z(z)P(z)=\frac{{\rm d}}{{\rm d}z}\left(D_z(z)P(z)\right)
\ .
\end{equation}
In the isotropic case, we have $\alpha=-1/2$ and $\gamma=3/4$. In this case we find
\begin{equation}
\label{eq: 4.1.15}
D_z=\frac{3}{8}(1-z^2)
\ ,\ \ \
v_z=-\frac{3}{4}z
\ ,\ \ \
{\rm (isotropic\ case)}
\end{equation}
and the normalised solution is $P(z)=\frac{1}{2}$ for $-1\le z\le 1$.

In the general case, we find that $(1-z^2)$ is a factor of $v_z-D_z'$, 
and the differential equation (\ref{eq: 4.1.14}) is 
\begin{equation}
\label{eq: 4.1.16}
\frac{1}{P}\frac{{\rm d}P}{{\rm d}z}=\frac{-z\left[6(5+4\alpha-4\gamma)z^2-13-11\alpha+10\gamma+4\Delta \mu \right]}{4\left[\gamma+(5+4\alpha-4\gamma)z^2-(5+4\alpha-4\gamma)z^4\right]}
\end{equation}
 it us useful to change the variable to $u=z^2$. In terms
of $u$, the differential equation (\ref{eq: 4.1.16}) may be written
\begin{equation}
\label{eq: 4.1.17}
\frac{1}{P}\frac{{\rm d}P}{{\rm d}u}=-\frac{6(5+4\alpha-4\gamma)u-13-11\alpha+10\gamma+4\Delta \mu}{8\left[\gamma+(5+4\alpha-4\gamma)u-(5+4\alpha-4\gamma)u^2\right]}
\ .
\end{equation}
Representing the right-hand-side using partial fractions, we obtain
\begin{equation}
\label{eq: 4.1.18}
\frac{1}{P}\frac{{\rm d}P}{{\rm d}u}=\frac{c_+}{u_+-u}+\frac{c_-}{u-u_-}
\end{equation}
where $u_\pm$ are the roots of the denominator on the right-hand-side of (\ref{eq: 4.1.17})
\begin{equation}
\label{eq: 4.1.19}
u_\pm=\frac{1}{2}\pm\frac{1}{2}\sqrt{1+\frac{4\gamma}{5+4\alpha-4\gamma}}
\end{equation}
and where the coefficients are
\begin{equation}
\label{eq: 4.1.20}
c_\pm=\frac{(4\Delta \mu- 2\alpha+\gamma-2)u_\pm-13-11\alpha-2\gamma-4\Delta \mu}
{4(5+4\alpha)}
\ .
\end{equation}
The probability density expressed in terms of $z$ is then
\begin{equation}
\label{eq: 4.1.21}
P(z)={\cal C}\,(z^2-u_-)^{c_-}(z^2-u_+)^{c_+}
\end{equation}
where ${\cal C}$ is a normalisation constant.

\subsection{Solution of rod alignment model}
\label{sec: 4.2}

Now we apply the solution obtained in section \ref{sec: 4.1} to the model for alignment of microscopic rods, as developed in sections \ref{sec: 2} and \ref{sec: 3}.
In section \ref{sec: 2.2} we introduced the Ornstein-Uhlenbeck
model for a random, isotropic velocity gradient field. 
The theory in section \ref{sec: 3} made two assumptions. 
In section \ref{sec: 3.1} it was assumed that the vorticity 
varies slowly, and section \ref{sec: 3.2} made a further assumption that the
strain field is small. Let us consider the implications of these assumptions for the parameters
of the model. The assumption that the vorticity varies slowly implies that $\tau_{\rm v}$
is large compared to other timescales in the system of equations. The typical 
strain rate $|S|=\sqrt{\langle {\rm tr}({\bf S}^2)\rangle}$ and the correlation time $\tau_{\rm s}$ should 
satisfy $|S|\tau_{\rm s}\ll 1$. The solution of the Ornstein-Uhlenbeck process implies 
\begin{equation}
\label{eq: 4.2.1}
\langle {\rm tr}({\bf S}^2)\rangle =10 D_{\rm s}\tau_{\rm s}
\end{equation}
so that the criterion for the strain to be small is simply 
$D_{\rm s}\tau^3_{\rm s}\ll 1$. The angular velocity 
$\omega$ is related to the magnitude of the vorticity $\Omega$ by
$\Omega=2\omega$. The magnitude of the vorticity is estimated by 
$\langle \Omega^2\rangle=-\frac{1}{2}{\rm tr}\langle (\mbox{\boldmath$\Omega$}^2)\rangle 
=3D_{\rm v}\tau_{\rm v}$. The rotation rate $\omega$ has a Gaussian distribution, with variance 
\begin{equation}
\label{eq: 4.2.2}
\sigma^2=\langle \omega^2\rangle=\frac{3}{4}D_{\rm v}\tau_{\rm v}
\ .
\end{equation}
Because the Ornstein-Uhlenbeck
model has an exponential decay of correlations, given by equation (\ref{eq: 2.2.3}), 
the spectral intensity of the strain fluctuations is a Lorentzian function:
\begin{equation}
\label{eq: 4.2.3}
I(\nu)=\frac{1}{1+\nu^2\tau_{\rm s}^2}
\ .
\end{equation}

In order to apply the results in section \ref{sec: 4.1} we must  
specify the covariances of the fluctuations of the axisymmetric effective strain 
tensor. 
If, in accord with the notation of section \ref{sec: 3.3}, we normalise the variances so that $\langle S_{11}^2\rangle=1$, $\langle S_{11} S_{22}\rangle=\alpha$,
$\langle S_{12}^2\rangle=\beta$, $\langle S_{13}^2\rangle=\gamma$, the non-zero covariances and expectation 
values of $\delta \Sigma_{ij}$ are 
\begin{eqnarray}
\label{eq: 4.2.4}
\langle \delta \Sigma_{11}^2\rangle =\langle \delta \Sigma_{22}^2\rangle 
&=\delta t\left[I(0)\left(\frac{1}{2}+\frac{\alpha}{2}\right)+I(2\omega)\left(\frac{1}{4}-\frac{\alpha}{4}+\frac{\beta}{2}\right)\right]
\nonumber \\
\langle \delta\Sigma_{11}\delta \Sigma_{22}\rangle
&=\delta t\left[I(0)\left(\frac{1}{2}+\frac{\alpha}{2}\right)-I(2\omega)\left(\frac{1}{4}-\frac{\alpha}{4}+\frac{\beta}{2}\right)\right]
\nonumber \\
\langle \delta\Sigma_{12}^2\rangle
&=\delta t\left[I(2\omega)\left(\frac{1}{4}-\frac{\alpha}{4}+\frac{\beta}{2}\right)\right]
\nonumber \\
\langle \Sigma_{13}^2\rangle=\langle \Sigma_{23}^2\rangle 
&=\delta t I(\omega)\gamma 
\nonumber \\
\langle \delta \Sigma_{33}^2\rangle&=2\,\delta t\,I(0)\,(1+\alpha) 
\nonumber \\
\langle \delta \Sigma_{11}\rangle=\langle\delta \Sigma_{22}\rangle&=\delta t
\left[I(0)\left(\frac{1}{4}+\frac{\alpha}{4}\right)+I(\omega)\frac{\gamma}{2}
+I(2\omega)\left(\frac{1}{4}-\frac{\alpha}{4}+\frac{\beta}{2}\right)\right]
\nonumber \\
\langle \delta \Sigma_{33}\rangle&=\delta t\left[I(0)(1+\alpha)+I(\omega)\gamma\right]
\ .
\end{eqnarray}
We use the assumption that the original random strain field $S_{ij}$ is isotropic, 
so that the statistics of these elements satisfy (\ref{eq: 3.3.6}). 
Using (\ref{eq: 4.2.4}) we obtain
\begin{eqnarray}
\label{eq: 4.2.5}
\langle \delta \Sigma_{11}^2\rangle=\langle \delta \Sigma_{22}^2\rangle&=\frac{\delta t}{4}[1+3I(2\omega)]
\nonumber \\
\langle \delta \Sigma_{11}\delta \Sigma_{22}\rangle &=\frac{\delta t}{4}[1-3I(2\omega)]
\nonumber \\
\langle \delta \Sigma_{12}^2\rangle&=\frac{\delta t}{4}3I(2\omega)
\nonumber \\
\langle \delta\Sigma_{13}^2\rangle=\langle \delta \Sigma_{23}^2\rangle &=\frac{\delta t}{4}3I(\omega)
\nonumber \\
\langle \delta \Sigma_{33}^2\rangle&=\delta t
\nonumber \\
\langle \delta \Sigma_{11}\rangle=\langle \delta \Sigma_{22} \rangle&=\frac{\delta t}{8}\left[1+3I(\omega)+6I(2\omega)\right]
\nonumber \\
\langle \delta \Sigma_{33}\rangle&=\frac{\delta t}{4}\left[2+3I(\omega)\right]
\ .
\end{eqnarray}
Normalising these by dividing by $\langle\delta \Sigma_{11}^2\rangle$, 
the anisotropy of the strain field induced by the vorticity is characterised
by modified forms for the parameters defined in (\ref{eq: 3.3.3}) and a scaled 
value of the parameter $\Delta \mu$ appearing in (\ref{eq: 4.1.13}):
\begin{eqnarray}
\label{eq: 4.2.6}
\alpha' &=\frac{1-3I(2\omega)}{1+3I(2\omega)}
\nonumber \\
\beta' &=\frac{3I(2\omega)}{1+3I(2\omega)}
\nonumber \\
\gamma' &=\frac{3I(\omega)}{1+3I(2\omega)}
\nonumber \\
\Delta \mu'&=\frac{3}{2}\frac{1+I(\omega)-2I(2\omega)}{1+3I(2\omega)}
\ .
\end{eqnarray}
Note that $\alpha'+2\beta'=1$, as expected from (\ref{eq: 3.3.5}). Because 
$I(0)=1$, when the vorticity is zero, we have $\alpha'=-1/2$, $\beta'=\gamma'=3/4$,
so that we recover the statistics of an isotropic strain field. 
In the limit as $\omega\to \infty$ we expect $I(2\omega)\ll I(\omega)\ll 1$, so that to leading order we may
set $\alpha=1$, $\beta=0$, $\Delta \mu=\frac{3}{2}$ and regard $\gamma$ as a small parameter.

Let us consider how to evaluate the probability density function of $z={\bf n}\cdot{\bf e}_\omega$
in the limit where $\omega\tau_{\rm s}\gg 1$. The general expression for the probability density is
equation (\ref{eq: 4.1.21}). In this limiting case, we may approximate the coefficients
$\alpha$ and $\gamma$ by
\begin{equation}
\label{eq: 4.2.7}
\alpha \sim 1
\ ,\ \ \ \ 
\gamma \sim \frac{3}{1+\omega^2\tau_{\rm s}^2}
\ ,\ \ \ \ 
\Delta \mu \sim \frac{3}{2}
\ .
\end{equation}
When $\gamma \ll 1$, the poles $u_\pm$ of the probability density function and the 
coefficients $c_\pm$ in (\ref{eq: 4.1.21}) are approximated by
\begin{eqnarray}
\label{eq: 4.2.8}
u_-\sim-\frac{\gamma}{9} \quad &u_+\sim1+\frac{\gamma}{9}
\nonumber \\
c_-\sim -\frac{1}{2}   \quad &c_+\sim -1
\end{eqnarray}
so that the probability density function is approximated by
\begin{equation}
\label{eq: 4.2.9}
P_\omega (z)\sim {\cal C}\left(z^2+\frac{\gamma}{9}\right)^{-1/2}\left(1+\frac{\gamma}{9}-z^2\right)^{-1}
\end{equation}
where ${\cal C}$ is a normalisation constant, and where the subscript $\omega$ is a reminder that this 
distribution is evaluated for a fixed value of $\omega$. 

\begin{figure}[t]
\centerline{\includegraphics[width=15.0cm]{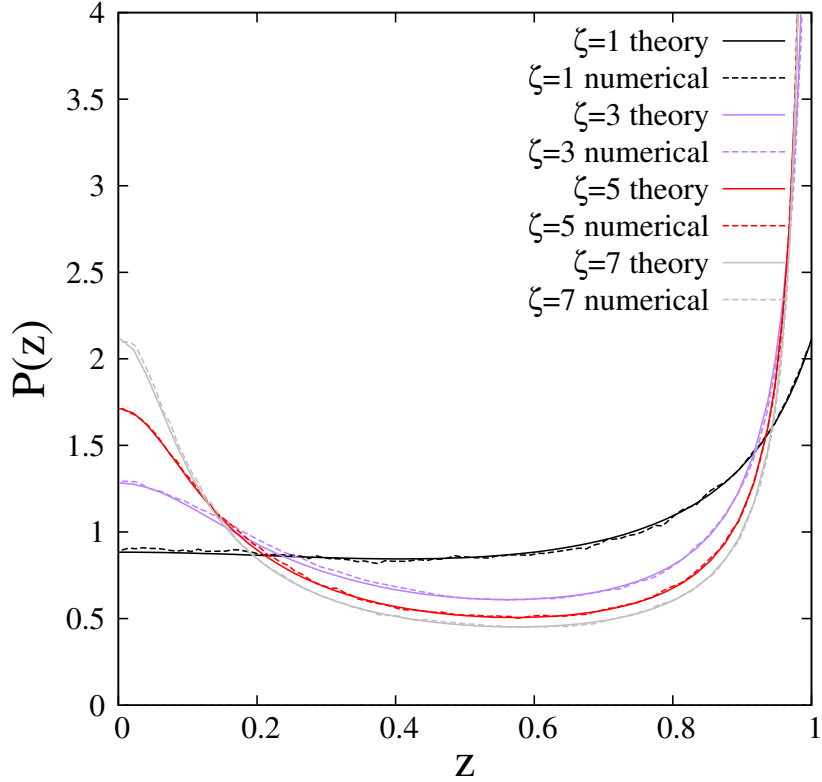}}
\caption{\label{fig: 1} 
Comparison between PDF of $z={\bf n}\cdot {\bf e}_3$ obtained by
simulation of Jeffery's equation of motion for the random strain model,
and the theoretical prediction, equations (\ref{eq: 4.1.19})-(\ref{eq: 4.1.21}) and (\ref{eq: 4.2.6}). In these simulations the vorticity is frozen 
so that $\zeta=\omega\tau_{\rm s}=1,3,5,7$.}
\end{figure}

We verified this relation by simulating the orientation of rod-like particles using equations (\ref{eq: 2.1.4}) 
and (\ref{eq: 2.1.5}), using the Ornstein-Uhlenbeck model for the velocity gradients. The components
of the vorticity were frozen, so that the only non-zero elements are $\Omega_{12}=-\Omega_{21}\equiv \omega$. 
The PDF of $z={\bf n}\cdot {\bf e}_3$ is plotted in figure \ref{fig: 1} for 
$\zeta=\omega\tau_{\rm s}=1,3,5,7$, showing good agreement with the theoretical 
prediction, equations (\ref{eq: 4.2.6}) and (\ref{eq: 4.1.19})-(\ref{eq: 4.1.21}). 
 The numerical simulations used $\tau_{\rm s}=1$, $D_{\rm s}=10^{-2}$, and the timestep
of the numerical integration was $dt=10^{-5}$ or smaller.

\begin{figure}[t]
\centerline{\includegraphics[width=15.0cm]{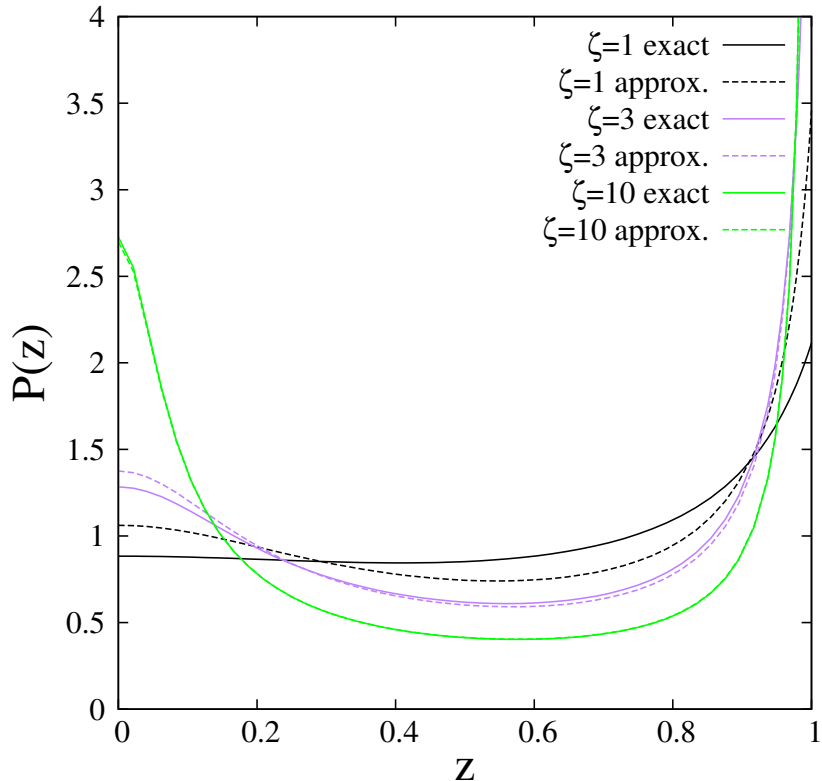}}
\caption{\label{fig: 2} 
Theoretical PDF of $z={\bf n}\cdot {\bf e}_3$ for the model where
the vorticity is frozen, with magnitude $\Omega=2\omega$. The distribution
(\ref{eq: 4.2.9}) is plotted for three values
of $\zeta=\omega\tau_{\rm s}$, namely $\zeta=1,3,10$. By comparison we have also 
plotted the exact probability density obtained from equations (\ref{eq: 4.1.19})-(\ref{eq: 4.1.21}) and (\ref{eq: 4.2.6}). At $\zeta=10$ the different plots are indistinguishable. }
\end{figure}

\begin{figure}[t]
\centerline{\includegraphics[width=15.0cm]{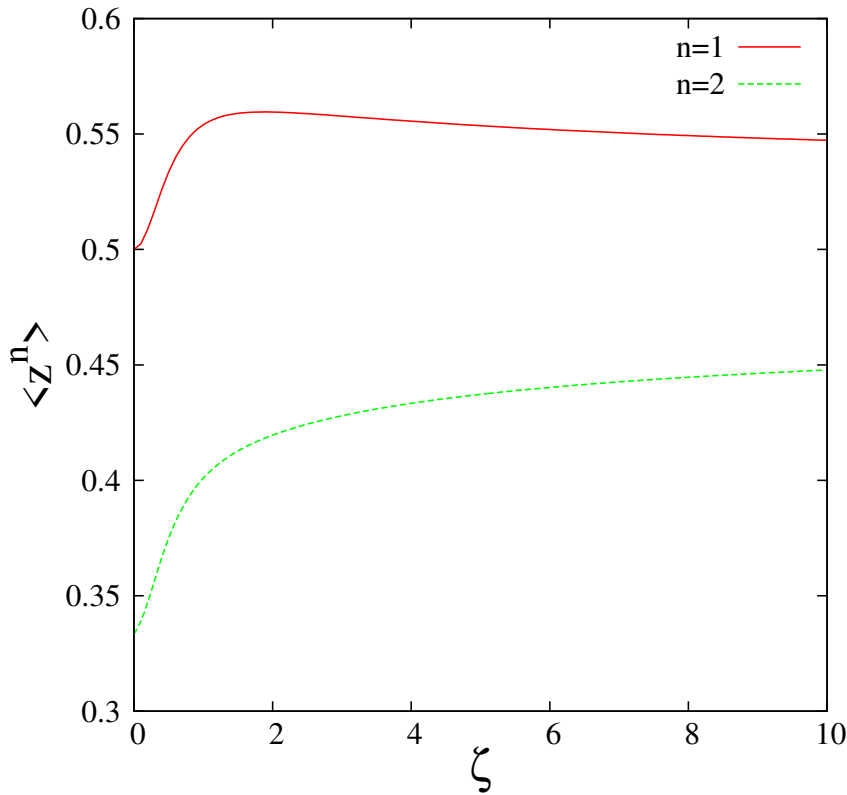}}
\caption{\label{fig: 3} 
Moments $\langle |z|\rangle$ and $\langle z^2\rangle$ as a function of $\zeta=\omega\tau_{\rm s}$.}
\end{figure}

In figure \ref{fig: 2} we plot the theoretical PDF of $z$ for three different values of the  
dimensionless variable $\zeta=\omega\tau_{\rm s}$, comparing (\ref{eq: 4.2.9}) with 
the exact expression obtained from using (\ref{eq: 4.2.6}) 
in equations (\ref{eq: 4.1.19}) -(\ref{eq: 4.1.21}). For $\omega\tau_{\rm s}=10$ 
plots of the exact and approximate PDF lie on top of each other. 
We observe that as $\omega\tau_{\rm s}\to \infty$
the distribution becomes concentrated around $z=\pm 1$ (rods perfectly aligned with the vorticity) 
and around $z=0$ (rods aligned perfectly perpendicular to the vorticity vector). The peak at $z=\pm 1$ is seen to be higher but narrower. In figure \ref{fig: 3} we plot
$\langle |z|\rangle$ and $\langle z^2\rangle$. Both of these statistics
approach $\frac{1}{2}$ in the limit as $\zeta\to \infty$, indicating that
in this limit both peaks carry half of the probability.

\begin{figure}[t]
\centerline{\includegraphics[width=15.0cm]{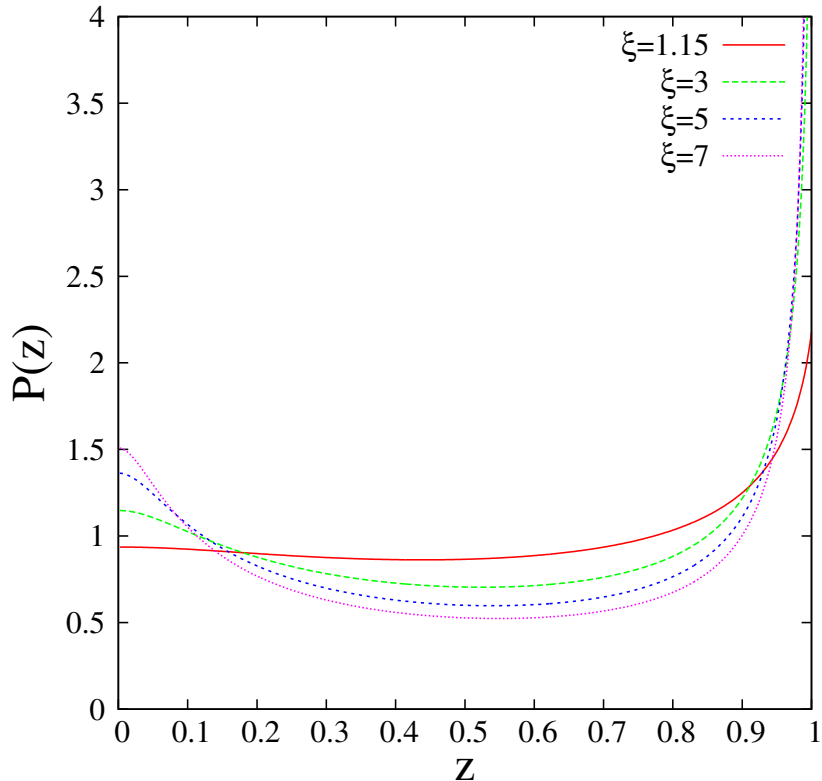}}
\caption{\label{fig: 4} 
Theoretical PDF of $z={\bf n}\cdot {\bf e}_3$, averaging over the 
slowly varying vorticity parameter $\omega$, which is Gaussian 
distributed with variance $\sigma^2=\frac{3}{4}D_{\rm v}\tau_{\rm v}$. 
The distribution (\ref{eq: 4.2.7}) is plotted for four values
of $\xi=\sigma\tau_{\rm s}$, namely $\xi=1.15,3,5,7$.}
\end{figure}

In practice the magnitude of the vorticity, $\omega$, is not frozen but fluctuates
slowly. It has a Gaussian distribution, with a variance 
$\sigma^2\equiv \langle \omega^2\rangle=\frac{3}{4}D_{\rm v}\tau_{\rm v}$. Our final
estimate for the probability density of $z$ is, therefore, the result on integrating the normalised 
PDF given by (\ref{eq: 4.2.9}) over a Gaussian distribution of $\omega$:
\begin{equation}
\label{eq: 4.2.10}
P(z)=\frac{2}{\sqrt{2\pi}\sigma}\int_0^\infty {\rm d}\omega\ \exp(-\omega^2/2\sigma^2)P_\omega(z)
\ .
\end{equation}
This PDF depends upon a single dimensionless parameter $\xi=\sigma\tau_{\rm s}$. The functions
obtained by numerical evaluation of the integral in (\ref{eq: 4.2.10}) are plotted in figure \ref{fig: 4} for three different values of $\xi=\sigma\tau_{\rm s}$. We used the exact
formulae for $P_\omega(z)$, equations (\ref{eq: 4.1.19})-(\ref{eq: 4.1.21}) and (\ref{eq: 4.2.6}), because
the integral includes the region where $\omega \tau_{\rm s}$ is small.

\section{Discussion}
\label{sec: 5}

We have determined the distribution of $z={\bf n}\cdot {\bf e}_\omega$ analytically for a 
model of microscopic rods in a random velocity field with isotropic statistics. The PDF 
shows a maximum at $z=1$ corresponding to alignment parallel to the vorticity,  
similar to findings of DNS studies of Navier-Stokes turbulence \cite{Pum+11}.

We conclude by making a few remarks about the relationship between the regime which
we have analysed and the velocity gradient statistics for Navier-Stokes turbulence.
The model for the velocity gradient of an isotropic flow which was introduced in section \ref{sec: 2.2}  
has four parameters, namely $\tau_{\rm v}$, $\tau_{\rm s}$, $D_{\rm v}$ and $D_{\rm s}$,
all of which have dimensions which depend only upon time. There are, therefore, three 
dimensionless parameters. In our analysis the vorticity was frozen,
so that $\tau_{\rm v}\to \infty$. The magnitude of the vorticity, which is of order 
$\omega\sim \sqrt{D_{\rm v}\tau_{\rm v}}$ was chosen so that $\zeta=\omega\tau_{\rm s}$ is finite. The 
diffusion coefficient $D_{\rm s}$ was assumed to be very small, so that the fluctuations 
of the strain are very small and may be treated using a Fokker-Planck equation.

In fact the form of the Navier-Stokes equation restricts the choice of parameters
in the Ornstein-Uhlenbeck model for the velocity gradient: it is well known that 
${\rm tr}(\mbox{\boldmath$\Omega$}^2)+{\rm tr}({\bf S}^2)=0$ \cite{Bet56},
which gives a further relation between $D_{\rm s}$ and $D_{\rm v}$. The 
Navier-Stokes equation also implies that the rate of dissipation per unit mass is 
${\cal E}=\nu{\rm tr}({\bf A}^{\rm T}{\bf A})$, which enables 
the norm of the velocity gradient to be expressed in terms of the Kolmogorov time,  
$\tau_{\rm K}=\sqrt{\nu/{\cal E}}$. These results imply the following relations, which 
determine the ratio of the diffusion coefficients $D_{\rm s}$ and $D_{\rm v}$ (see \cite{Pum+11}):
\begin{equation}
\label{eq: 5.1}
D_{\rm s}=\frac{1}{20\tau_{\rm K}^2\tau_{\rm s}}
\ ,\ \ \
D_{\rm v}=\frac{1}{12\tau^2_{\rm K}\tau_{\rm v}}
\ .
\end{equation}
Numerical studies indicate that the exponential correlation function is a reasonable 
model for the statistics of fully developed turbulence, with the parameters 
$\tau_{\rm s}$ and $\tau_{\rm v}$ satisfying $\tau_{\rm s}\approx 2.3 \tau_{\rm K}$
and $\tau_{\rm v}\approx 7.2 \tau_{\rm K}$ (these are the values quoted in \cite{Pum+11},
which discusses earlier work on the velocity gradient correlation functions). This justifies 
the assumption that the vorticity is slowly varying, and the variance of the vorticity parameter is estimated
to be $\sigma^2=\langle \omega^2\rangle =\frac{3}{4}D_{\rm v}\tau_{\rm v}$, 
so $\sigma= 1/(2\tau_{\rm K})$, implying that
$\xi=\sigma \tau_{\rm s}\approx 1.15$ is the value which should be compared with
the data on alignment in Navier Stokes turbulence, discussed in \cite{Pum+11}. 
There is a qualitative but not quantitative agreement between the curve in figure \ref{fig: 4} for $\xi=1.15$
and the results of DNS simulations in figure 2 of \cite{Pum+11}: both show a peak
in the PDF at $z=1$, but this peak is more pronounced in the DNS data.
We conclude that our model should be understood as a laboratory 
for understanding alignment of microscopic rods with vorticity, rather than providing
a quantitative description.

{\sl Acknowledgements}. HRK thanks the Open University for a postgraduate studentship.

\end{document}